\documentclass[twocolumn]{revtex4}
\usepackage{graphicx}
\usepackage{dcolumn}
\usepackage{subfigure}
\usepackage{bm}

\begin{document}

\title{Transition Voltage Spectroscopy and the nature of vacuum tunneling}

\author{M.L. Trouwborst}
\email{Trouwborst@physics.Leidenuniv.nl}
\affiliation{Kamerlingh Onnes Laboratorium, Leiden University, P.O. Box 9504, 2300 RA Leiden, The Netherlands}
\author{C.A. Martin}
\affiliation{Kamerlingh Onnes Laboratorium, Leiden University, P.O. Box 9504, 2300 RA Leiden, The Netherlands}
\affiliation{Kavli Institute of Nanoscience, Delft University of Technology, P.O. Box 5046, 2600 Ga Delft, The Netherlands}
\author{R.H.M. Smit}
\affiliation{Kamerlingh Onnes Laboratorium, Leiden University, P.O. Box 9504, 2300 RA Leiden, The Netherlands}
\author{C.M. Gu\'{e}don}
\affiliation{Kamerlingh Onnes Laboratorium, Leiden University, P.O. Box 9504, 2300 RA Leiden, The Netherlands}
\author{T.A. Baart}
\affiliation{Kamerlingh Onnes Laboratorium, Leiden University, P.O. Box 9504, 2300 RA Leiden, The Netherlands}
\author{S.J. van der Molen}
\affiliation{Kamerlingh Onnes Laboratorium, Leiden University, P.O. Box 9504, 2300 RA Leiden, The Netherlands}
\author{J.M. van Ruitenbeek}
\affiliation{Kamerlingh Onnes Laboratorium, Leiden University, P.O. Box 9504, 2300 RA Leiden, The Netherlands}

%%%%%%%%%%%%%%%%%%%%%%%%%%%%%%%%%%%%%%%%%%%%%%%%%%%%%%%%%%%%%%%%%%%%%
%% The document title should be given as usual
%% A short title can be given as a *suggestion* for running headers.
%%%%%%%%%%%%%%%%%%%%%%%%%%%%%%%%%%%%%%%%%%%%%%%%%%%%%%%%%%%%%%%%%%%%%

\begin{abstract}
Transition Voltage Spectroscopy (TVS) has been proposed as a tool to
analyze charge transport through molecular junctions. We extend TVS to
Au-vacuum-Au junctions and study the distance dependence of the
transition voltage $V_{\rm t}(d)$ for clean electrodes in cryogenic vacuum. On the one hand, this
allows us to provide an important reference for $V_{\rm t}(d)$-measurements on
molecular junctions. On the other hand, we show that TVS forms a simple
and powerful test for vacuum tunneling models.
\end{abstract}

\maketitle

Molecular electronics offers the exciting prospect of tuning the conductance of nanometer-sized junctions with the tools of synthetic chemistry. In principle, it is possible to fabricate small switchable devices from functional molecules \cite{switches}. However, the electronic properties of a molecule change dramatically when it is contacted by metal electrodes, as the coupling significantly shifts and broadens the molecular levels \cite{switches,liljeroth,dulic}. Hence, it is of utmost importance to understand the influence of the electrodes on the molecular levels. The level structure can in principle be studied by measuring the current-voltage (I-V) characteristics of molecular junctions; as soon as a molecular level becomes accessible by the bias window, a step in the current can be observed. However, such experiments are complicated by the fact that the position of the molecular level is often a few eV away from the Fermi energy $E_F$. As a consequence, very large electric fields need to be applied to probe these levels, often resulting in a breakdown of the junction before the first molecular level can be accessed.

Recently, Beebe \textit{et al.} have presented a new method to characterize the electronic structure of molecular junctions which requires lower electric fields \cite{beebe,beebe_acs}. By plotting their I-V data in a Fowler-Nordheim (FN) representation, i.e. plotting $\ln(I/V^2$) versus $1/V$ \cite{FN}, they observed a minimum at a characteristic voltage $V_{\rm t}$. Interestingly, for a broad set of molecules $V_{\rm t}$ was found to be proportional to the position of the highest occupied molecular orbital ($\epsilon_{H}$) with respect to $E_F$, as confirmed by ultraviolet photoelectron spectroscopy (UPS). Hence, TVS was proposed as a new, simple and powerful spectroscopic tool to study the positions of molecular levels.

However, the interpretation of TVS is still under debate \cite{huisman,troels,araidai}. In the original picture of Beebe \textit{et al.} a molecular junction was treated as a simple tunnel junction. They interpreted $V_{\rm t}$ as the potential for which the tunnel barrier, tilted by the bias voltage, changes from trapezoidal to triangular. In this picture, $V_{\rm t}$ would be the same for all electrode separations $d$. In contrast, Huisman \textit{et al.} considered the tunnel barrier model in more detail and calculated that $V_{\rm t}$ should be strongly dependent on $d$ \cite{huisman,simmons,stratton}. They obtained the following approximation for $V_{\rm t}$ in the case of a trapezoidal vacuum barrier with height $\phi$ (m is the electron mass):

\begin{equation}\label{Vtrans}
   V_{\rm{t}} \approx \frac{2\hbar}{e\sqrt{m}}\frac{\sqrt{2\phi}}{d}
\end{equation}

Remarkably, such a variation of $V_{\rm t}$ with $d$ is much stronger than the weak distance dependence reported for junctions containing aliphatic and conjugated molecules of different lengths \cite{beebe_acs}. This statement remains valid even when including image forces which modify the potential barrier \cite{huisman}. Therefore, Huisman \textit{et al.} suggested that the dependence $V_{\rm t}({d})$ might form a new tool to distinguish molecular junctions from vacuum junctions. However, TVS studies on vacuum junctions have not been reported thus far.

To test this proposition we present extensive TVS measurements on metal-vacuum-metal junctions. The possibility to accurately vary the tunnel gap between the electrodes allows us to reveal the basic properties of TVS. We show that for Au junctions $V_{\rm t}$ only has a weak variation with electrode spacing $d$, comparable to the dependence reported for molecular junctions. Hence, our results indicate that the $V_{\rm t}(d)$ dependence itself cannot be used to distinguish molecular junctions from vacuum junctions. However, the absolute values of $V_{\rm t}$ are significantly higher for vacuum barriers than for junctions with conjugated molecules. Interestingly, our study demonstrates that TVS can also be used as a characterization tool for vacuum barriers. The dependence $\phi(d)$ in standard models for vacuum tunneling, such as those leading to Eq. \ref{Vtrans}, can be tested by TVS measurements.

%\paragraph{ EXPERIMENTAL DETAILS }

\begin{figure}[ht]
\begin{center}
\includegraphics[width=8.7cm]{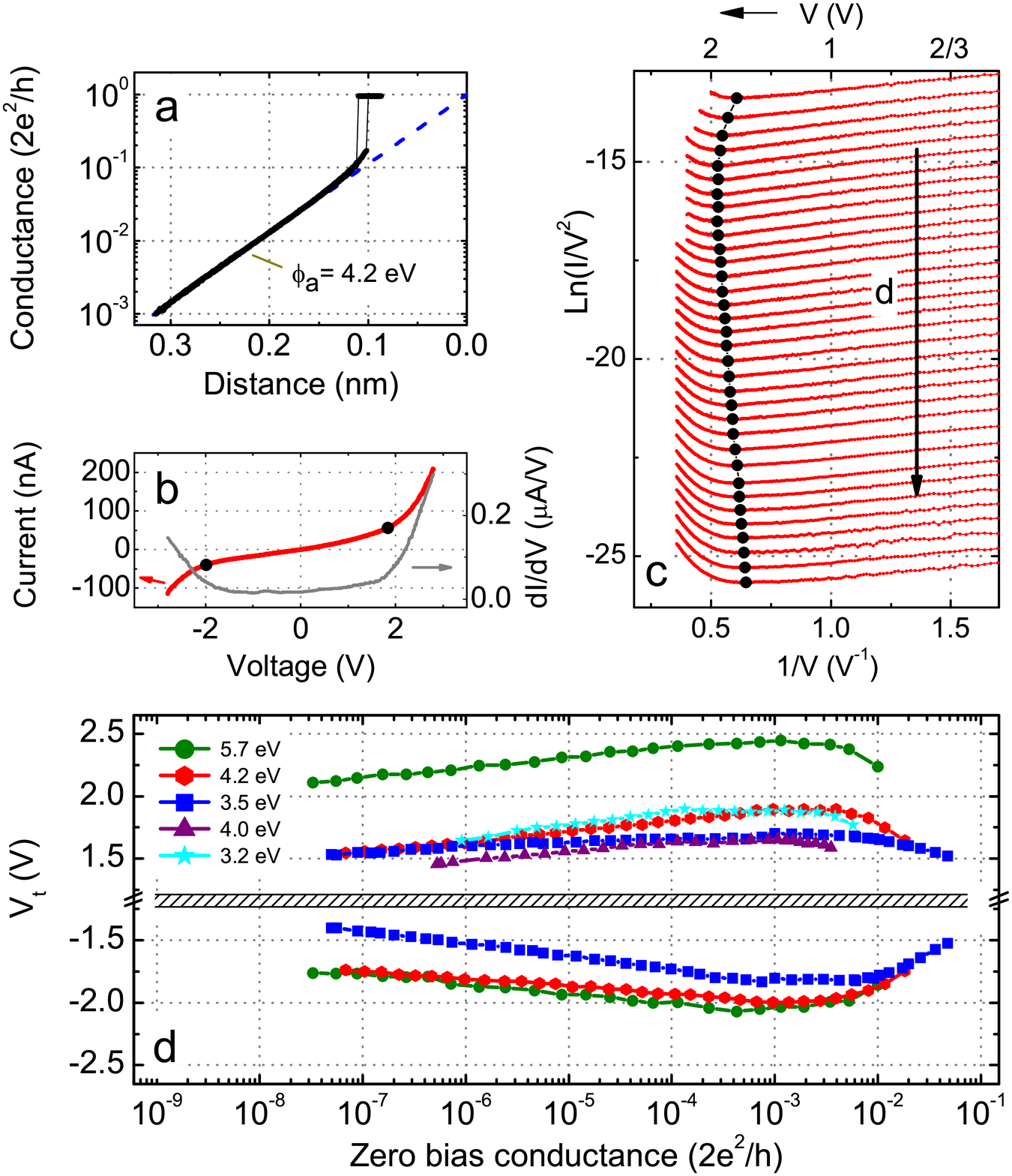}
\end{center}
\caption{Electrical characteristics of clean gold junctions at T$\approx$ 5 K. a) Conductance versus width of the vacuum gap of a trained junction at a bias voltage of $V=100$ mV. From the exponential decay at large distance, an apparent barrier height is deduced of $4.2\pm1$ eV (Eq. \ref{App_barrier}) \cite{chains,calibratie2}. b) Typical I-V curve in the tunneling regime. Here, $V_{\rm t}$ is -1.99 V for negative voltage and 1.84 V for positive voltage (black dots), as obtained from: c) Fowler-Nordheim plot of the I-V characteristics for 34 different positions. After each curve the electrode separation is increased by $0.02$ nm, resulting in a lower current (no offset is used). The black dots represent the minima, or $V_{\rm t}$. Remarkably, $V_{\rm t}$ decreases with distance for wide tunnel barriers while it increases for short tunnel barriers. d) $V_{\rm t}$ versus zero bias conductance for different contacts, measured on 3 different samples. Note the break in the scale between -1.2 V and +1.2V. The two curves marked by triangles and stars are measured on the same sample, but the latter was obtained after modifying the electrodes. The same holds for the two curves marked by squares and hexagons. For each contact, the apparent barrier height was measured and its value is given next to the data points ($\pm$ 1 eV).}
\label{TVS_overzicht}
\end{figure}

For a full characterization of TVS on a metal-vacuum-metal junction, voltages up to 3 V are to be applied over vacuum gaps as small as $\approx$0.3 nm. Such high electric fields (and high field gradients) may cause instabilities in the tunnel junctions. Hence, junctions are needed which are stable in time and kept in a clean environment. For this reason we used notched-wire mechanically controllable break junctions in cryogenic vacuum (T$\approx$ 5 K) \cite{mcbj,review}. The electrodes are made of gold, the archetypical electrode metal for molecular junctions. In addition, the junctions were first optimized by a "training" procedure, i.e. by repeatedly opening and closing the electrodes \cite{PRL_Marius}. We expect that this organizes the apex atoms into their strongest bond configuration and enhances their stability in high electric fields. The high stability and repeatability of the conductance evolution is illustrated by the tunnel curve in Fig. \ref{TVS_overzicht}a. Upon closing, the tunnel current increases exponentially until the electrodes snap to contact \cite{PRL_Marius,untiedt}. Note that the conductance jumps to a value close to 1 $G_0$ (1$G_0 = 2e^2/h$), indicating a clean single-atom contact. After the training procedure, the electrodes are separated such that a vacuum gap is created with a zero bias conductance of $\approx$0.01 $G_0$. This is the starting point for the TVS measurements. Subsequently, the vacuum gap is increased stepwise, and an I-V curve over the range $\pm$2-3 V is recorded for each position.

A typical example of such an I-V curve is plotted in Fig. \ref{TVS_overzicht}b. Clearly, the current displays a transition from a linear dependence at low voltages to a strongly nonlinear behavior for voltages $>1.5 V$. As stated earlier, this transition can be quantified by scaling the data in a Fowler-Nordheim representation. This is shown in Fig. \ref{TVS_overzicht}c, here for positive bias voltage only. Let us first discuss the upper curve. This curve is measured for a small tunnel gap with a zero bias conductance of $\approx$0.02 $G_0$. It has a well-defined minimum at 1.64 V that determines $V_{\rm t}$. In total, 34 curves are plotted, corresponding to 34 different electrode separations (equally spaced by $\approx$0.02 nm).  When we increase the electrode separation, a shift in $V_{\rm t}$ can be observed. The transition voltage first increases with distance and is at a maximum after stretching by $\approx$0.1 nm (fifth curve). $V_{\rm t}$ has now increased to 1.9 V. For even larger gaps $V_{\rm t}$ decreases again to a value of 1.55 V after stretching by $\approx$0.6 nm.

Before we continue our discussion on the dependence of $V_{\rm t}$ on distance, we first study the reproducibility of the measurements. We have repeated the experiments on different contacts and for 3 samples. All data are plotted in Fig. \ref{TVS_overzicht}d. For all junctions $V_{\rm t}$ has a similar dependence on distance, and $V_{\rm t}$ has a maximum around $10^{-3}$ $G_0$. However, there is a strong variation in the absolute value of $V_{\rm t}$. For example around $10^{-7}$ $G_0$, $V_{\rm t}$ scatters between 1.4 and 2.2 V. Similarly, $V_{\rm t}$ usually differs for positive and negative voltage polarity in the same contact. This suggests that $V_{\rm t}$ is sensitive to the local shape of the electrodes. This is confirmed by closing the electrodes to a conductance $>10 G_0$ before starting a new measurement cycle. After this modification of the electrodes, $V_{\rm t}$ was found to have a different value. This is illustrated by the filled triangles and filled stars in Fig. \ref{TVS_overzicht}d which represent two data sets that were obtained before and after such a reconfiguration. Actually, modifying the electrodes also changes the work function. This can be concluded by extracting the apparent barrier heights from the slopes of the exponential decay of the tunneling conductance for the individual contacts at large $d$ (see Fig. \ref{TVS_overzicht}a). The apparent barrier height is defined as

\begin{equation}\label{App_barrier}
   \phi_a=(\hbar \partial \ln G / \partial d)^2/8m
\end{equation}

For low voltages, $\phi_a$ is expected to be almost equal to the work function of the electrodes \cite{blanco,binnig}. For our junctions $\phi_a$ was found to vary between 3.2 en 5.7 eV, which is in the range of typical values for gold electrodes with different crystal orientations \cite{olesen}. Remarkably, the contact corresponding to the green bullets in Fig. \ref{TVS_overzicht}d shows the highest values for both $V_{\rm t}$ and $\phi_{a}$. This brings us to the first conclusion: to obtain a reliable value of $V_{\rm t}$ the variation of the atomic configuration of the electrodes needs to be taken into account.

\begin{figure}[ht]
\begin{center}
\includegraphics[width=8cm]{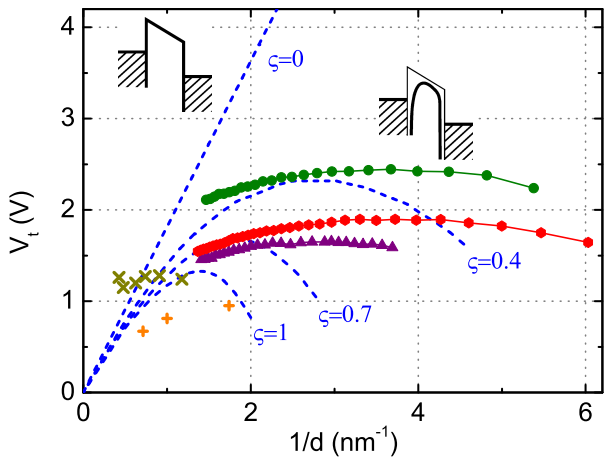}
\end{center}
\caption{Variation of the transition voltage with electrode distance. Filled symbols: measurements of $V_{\rm t}$ vs $1/d$ for 3 different gold samples in vacuum. Dashed straight line: $V_{\rm t}$ vs $1/d$ as deduced from Eq. \ref{Vtrans} ($\phi=$5.4 eV). Dashed curved lines: $V_{\rm t}$ vs $1/d$ as calculated from the full Simmons model including the image forces. The strength of the image forces is varied by changing the prefactor $\zeta$ in Eq. \ref{Vtrans} from 0.4 to 1. "$\times$" and "+": data from large-area molecular junctions as reported by Beebe \textit{et al.} "$\times$": Alkane series, C$_{18}$-SH, C$_{16}$-SH, C$_{12}$-SH, C$_{10}$-SH, C$_{8}$-SH, and C$_{6}$-SH. "+": phenyl series, TP-SH, BP-SH, and Ph-SH \cite{beebe_acs}.}
\label{Model}
\end{figure}

Let us now return to the distance dependence of $V_{\rm t}$. In order to directly compare our measurements to the predictions of Eq. \ref{Vtrans}, we need to plot the data as a function of $1/d$. For this purpose, the origin in the position $(d=0)$ was defined by extrapolating the exponential part of Fig. \ref{TVS_overzicht}a (dashed line) to a conductance of $2e^{2}/h$. The crossing point is then set as the origin. As a result we obtain Fig. \ref{Model}. There is a striking difference when comparing the experimental data and the straight line expected from Eq. \ref{Vtrans}: our data are not proportional to $1/d$. Instead, only a modest variation with $d$ is found, with a maximum at $\approx3^{-1}$ $nm^{-1}$. Clearly, the square barrier model (with a constant height $\phi$) does not give an accurate description of the data.

As discussed by Simmons and applied by Huisman, a more realistic representation of the barrier takes image forces into account \cite{huisman,simmons}. Image forces effectively round and lower the tunnel barrier. This can be described by the following equation for the shape of the barrier at a position x from the electrode surface:
\begin{equation}\label{image}
    \phi(x)=\phi_0-eV\frac{x}{d}-0.58 \zeta \frac{e^2 \ln2}{4 \pi \epsilon} \frac{d}{x(d-x)}
\end{equation}
where $\zeta$ is a variable we inserted to fit the strength of the image force \cite{huisman,simmons}. Actually, we note that Eq. \ref{image} not only applies for image forces; in general, it may be regarded as a more elaborate description of the barrier. For example, one could also use a $1/r$ potential to describe the effective ion potentials  of the outermost atoms in the two electrodes. At zero bias, the potentials of the opposing atoms will add up as $-1/x - 1/(d-x))$ to $-d/(x(d-x))$, resulting in the same type of expression as Eq. \ref{image}.

Using the barriers described by Eq. \ref{image}, a different relation is obtained for $V_{\rm t}(d)$, which is plotted in Fig. \ref{Model}. Interestingly, Eq. \ref{image} yields a maximum in $V_{\rm t}$ at a certain distance, in correspondence with the experimental data. We find the best qualitative agreement between the calculated values of $V_{\rm t}$ and the experimental values for a reduced value of the image force $\zeta=0.4-0.7$. This can be expected since the image forces for atomically sharp electrodes are supposedly not as large as for infinitely large flat plates (as used by Simmons). Overall, the rounded barrier gives a better description of the data. However, the variation of the experimental data with $d$ is still much less pronounced than in the models.

Let us first discuss the implications of our data for the interpretation of TVS. As described before, we want to make a comparison between the distance dependence of $V_{\rm t}$ for vacuum junctions and molecular junctions. For this purpose, we also include two data sets measured by Beebe \textit{et al.} in Fig. \ref{Model} \cite{beebe_acs}. The upper data set ("$\times$"), corresponding to alkanethiols of different lengths, has a negligible variation in  $V_{\rm t}$. This was ascribed to an almost constant HOMO-LUMO gap for different alkane lengths. The lower data set ("+") corresponds to ($\pi -$conjugated) phenylene molecules. Compared to the alkanethiols, these molecules have a stronger dependence of $V_{\rm t}$ on $d$. This was attributed to a variation in the HOMO-LUMO gap, which is expected to decrease with increasing molecule length. Let us now compare the measurements. The two types of data (molecular junctions and vacuum junctions) have been measured at slightly different distances. In contrast to our MCBJ measurements, the experiments of Beebe \textit{et al.} were carried out on large-area junctions, in which many molecules are probed in parallel. In addition, the distance between the Fermi level and the nearest molecular level is lower than the work function of the electrodes. As a result, the conductance is larger for the large-area molecular junctions which makes it possible to measure at larger distances or smaller values of $1/d$, respectively. Nevertheless, we find that the distance dependence of $V_{\rm t}$ for the molecular data does not differ significantly from that observed in our vacuum measurements. This is an important conclusion of this paper. Taking into account (i) the measurement accuracy of the molecular data of approximately $\pm100$ mV and (ii) the limited variation of $V_{\rm t}$ with $d$ for the vacuum data it is not possible to distinguish molecular junctions from vacuum junctions just by measuring the distance dependence of $V_{\rm t}$. However, considering the absolute values of $V_{\rm t}$ for conjugated molecules there is a clear difference with the vacuum data. For conjugated molecules, the reported values for $V_{\rm t}$ are much lower (0.6 V to 1 V) than the values found for the vacuum junctions ($>1.4$ V) \cite{troels}. This difference can thus be regarded as a signature of molecular junctions.

Let us return to Fig. \ref{Model} and focus on the tunnel data only. Another way to interpret this graph, is that we have used TVS to test the tunnel barrier models described by Eqs. \ref{Vtrans} and \ref{image}. In principle, any type of tunnel model could be tested, making TVS a potential tool for more detailed studies on electron transmission in a variety of tunnel systems. This is the third conclusion of our study. In our specific case, TVS allows us to conclude that the models in Eqs. \ref{Vtrans} and \ref{image} are incomplete, as they do not predict the observed weak $V_{\rm t}(d)$-dependence.

We note that other physical mechanisms affecting $V_{\rm t}$ are electrostatic forces \cite{olesen,PRL_Marius,martin,metallic_adhesion} and thermal expansion. However, these can only lead to an additional lowering of $V_{\rm t}$ for very small gaps ($<0.25$ nm). Hence, we should consider other mechanisms to explain the relatively weak variation of $V_{\rm t}$ with $d$. Actually, there is also another indication that the theory behind TVS is not yet complete. More complex TVS transitions have recently  been reported in literature, where there is not a single well-defined minimum in the FN plot, but rather a plateau \cite{Choi,Lu}. These transitions have not been fully explained. In fact, we have also observed more complex transitions. In about 15$\%$ of the contacts, two minima appear close to each other (see supporting info). The fact that these complex structures appear for clean junctions without molecules indicates there are other physical mechanisms which could affect TVS. A possible explanation is due to the local density of states (LDOS) of the electrodes \cite{discussions}. If surface states are present, the LDOS of Au is strongly dependent on the electron energy. Thus, it is very well possible that a peak in the LDOS results in a minimum in the FN plot. Since the LDOS strongly depends on the crystal orientation and the exact shape of the electrodes, double minima may appear experimentally. Two minima generally only appear for one polarity of the voltage, indicating that the appearance of two minima is critically related to the exact properties of one of the two electrodes.

Summarizing, we have performed TVS measurements on metal-vacuum-metal junctions. From the vacuum measurements we have learned the following aspects for TVS. First, $V_{\rm t}$ is found to be sensitive to the local shape of the electrodes. Thus, for a reliable value of $V_{\rm t}$ one should obtain information on the detailed configuration, or otherwise average over many configurations. Second, $V_{\rm t}$ is found to be sensitive to the distance between the electrodes. However, the relation between $V_{\rm t}$ and electrode separation $d$ is much weaker than expected from standard models for vacuum tunneling and is comparable to the variation found for molecular junctions. Hence, it is not possible to use the variation of $V_{\rm t}$ on $d$ to distinguish molecular junctions from vacuum junctions, as suggested recently, but the value of $V_{\rm t}$ provides a strong indication. Finally, we propose TVS as a simple but powerful test for models of vacuum tunneling.

Acknowledgments. We thank Eek Huisman, Kristian Thygesen and Troels Markussen for inspiring discussions. The latter two are also acknowledged for suggesting the importance of surface states on the electrodes. This work is part of the research programme of the Foundation for Fundamental Research on Matter (FOM), which is part of the Netherlands Organisation for Scientific Research (NWO).

Supporting Information Available: The calibration method for the mechanically controllable break junctions and measurements displaying multiple minima close to each other.


\begin{thebibliography}{9}

\bibitem{switches}
S.J. van der Molen and P. Liljeroth, \textit{J. Phys.: Condens. Matter} \textbf{2010}, 22, 133001.

\bibitem{liljeroth}
P. Liljeroth, J. Repp, and G. Meyer, \textit{Science} \textbf{2007}, 317, 1203.

\bibitem{dulic}
D. Dulic, S.J. van der Molen, T. Kudernac, H.T. Jonkman, J.J.D. de Jong, T.N. Bowden, J. van Esch, B.L. Feringa, and B.J. van Wees, \textit{Phys. Rev. Lett.} \textbf{2003}, 91, 207402.

\bibitem{beebe}
J.M. Beebe, B. Kim, J.W. Gadzuk, C.D. Frisbie, J.G. Kushmerick, \textit{Phys. Rev. Lett.} \textbf{2006}, 97, 026801.

\bibitem{beebe_acs}
J.M. Beebe, B. Kim, C.D. Frisbie, J.G. Kushmerick, \textit{ACS Nano} \textbf{2008}, 2, 827.

\bibitem{FN}
R.H. Fowler, and L. Nordheim, \textit{Phys. R. Soc. A} \textbf{1928}, 119, 173.

\bibitem{huisman}
E.H. Huisman, C.M. Guedon, B.J. van Wees and S.J. van der Molen, \textit{Nano Lett.} \textbf{2009}, 9, 3909.

\bibitem{araidai}
M. Airaidai, and M. Tsukada, \textit{Phys. Rev. B} \textbf{2010}, 81, 235114.

\bibitem{troels}
J. Chen, T. Markussen, and K.S. Thygesen, \textit{Phys. Rev. B} \textbf{2010}, 82, 121412.

\bibitem{simmons}
J.G. Simmons, \textit{J. Appl. Phys.} \textbf{1963}, 34, 1793.

\bibitem{stratton}
R.J. Stratton, \textit{Phys. Chem. Solids} \textbf{1962}, 23, 1177.

\bibitem{blanco}
J.M. Blanco, F.N. Flores, R. Perez, \textit{Progress in Surface Science} \textbf{2006}, 81, 403.

\bibitem{binnig}
G. Binnig, N. Garcia, H. Rohrer, J.M. Soler and F. Flores, \textit{Phys. Rev. B.} \textbf{1984}, 30, 4816.

\bibitem{review}
N. Agra\"{\i}t, A. Levy-Yeyati and J.M. van Ruitenbeek, \textit{Phys. Rep.} \textbf{2003}, 377, 81.

\bibitem{mcbj}
J. Moreland and J.W. Ekin, \textit{J. Appl. Phys.} \textbf{1985}, 58, 3888.

\bibitem{PRL_Marius}
M.L. Trouwborst, E.H. Huisman, F.L. Bakker, S.J. van der Molen, and B.J. van Wees, \textit{Phys. Rev. Lett.} \textbf{2008}, 100, 175502.

\bibitem{untiedt}
C. Untiedt, M.J. Caturla, M.R. Calvo, J.J. Palacios, R.C. Segers, and J.M. van Ruitenbeek , \textit{Phys. Rev. Lett.} \textbf{2007}, 98, 206801.

\bibitem{chains}
A.I. Yanson, G. Rubio Bollinger, H.E. van den Brom, N. {Agra\"{\i}t} and J.M. van Ruitenbeek, \textit{Nature} \textbf{1998}, 395, 783.

\bibitem{calibratie2}
C. Untiedt, A.I. Yanson, R. Grande, G. Rubio-Bollinger, N. {Agra\"{\i}t}, S. Vieira, and J.M. van Ruitenbeek, \textit{Phys. Rev. B} \textbf{2002}, 66, 085418.

\bibitem{olesen}
L. Olesen, M. Brandbyge, M.R. S{\o}rensen, K.W. Jacobsen, E. L{\ae}gsgaard, I. Stensgaard, and F. Besenbacher, \textit{Phys. Rev. Lett.} \textbf{1996}, 76, 1485.

\bibitem{discussions}
T. Markussen, K. Thygesen, and K.W. Jacobsen, preprint.

\bibitem{martin}
C.A. Martin, R.H.M. Smit, H.S.J. van der Zant, and J.M. van Ruitenbeek, \textit{Nano Lett.} \textbf{2009}, 9, 2940-2945.

\bibitem{metallic_adhesion}
G. Rubio-Bollinger, P. Joyez and N. Agra\"{\i}t, \textit{Phys. Rev. Lett.} \textbf{2004}, 93, 116803.

\bibitem{Choi}
S. H. Choi, B. Kim, C. D. Frisbie, \textit{Science} \textbf{2008}, 320, 1482.

\bibitem{Lu}
Q. Lu, K. Liu, H.M. Zhang, Z.B. Du, X.H. Wang, F.S. Wang, \textit{ACS Nano} \textbf{2009}, 3 (12), 3861.

\end{thebibliography}
\end{document}